# Single-crystalline GaAs/Si Heterojunction Tunnel Diodes Interfaced by an Ultrathin Oxygen-enriched Layer

Jie Zhou, *Member, IEEE*, Yifan Wang, Ziqian Yao, Qingxiao Wang, Yara S. Banda, Jiarui Gong, *Member, IEEE*, Yang Liu, *Member, IEEE*, Carolina Adamo, Patrick Marshall, *Member, IEEE*, Yi Lu, Tsung-Han Tsai, Yiran Li, Vincent Gambin, Tien Khee Ng, Boon S. Ooi, *Fellow, IEEE*, and Zhenqiang Ma, *Fellow, IEEE*

*Abstract*—We report the fabrication and characteristics of GaAs/Si $p^+/n^+$ heterojunction tunnel diodes. These diodes were fabricated via grafting the freestanding single-crystalline p-type degenerately doped GaAs ($4 \times 10^{19}$ cm$^{-3}$) nanomembrane (NM) onto single-crystalline n-type Si ($5 \times 10^{19}$ cm$^{-3}$) substrate. At the heterointerface, an amorphous ultrathin oxygen-enriched layer (UOL) was intentionally engineered through chemical oxidation and atomic layer deposition (ALD). Scanning transmission electron microscopy (STEM) confirmed the formation of the UOL and the single crystallinity of the grafted junction. The resulting tunnel diodes consistently exhibited negative differential resistance (NDR) behavior at room temperature, with a high maximum peak-to-valley current ratio (PVCR) of 36.38, valley voltages ranging from 1.3 to 1.8 V, and a peak tunneling current density of 0.95 kA/cm$^2$. This study not only highlights the critical roles of the UOL as both an interface improvement layer and a quantum tunneling medium, but also establishes "semiconductor grafting" as an effective and versatile method for high-performance, lattice-mismatched heterojunction devices.

## I. INTRODUCTION

LATTICE-MATCHED semiconductor heterojunction is the key building block of modern electronic and optoelectronic devices [1], [2], [3]. However, the formation of heterojunctions between dissimilar semiconductors, like GaAs and Si, through conventional techniques including heteroepitaxy and direct bonding, faces significant challenges due to lattice, and thermal-mechanical mismatches [1]. Such mismatches can introduce interfacial defects and dislocations, delamination, and low bonding yield, *etc.*, thus degrading device performance and reliability.

GaAs/Si heterojunctions have attracted considerable interest due to the complementary properties and unique band alignment of GaAs and Si. While GaAs offers a direct bandgap and high electron mobility, Si provides a robust and scalable integration platform. A wide range of GaAs/Si devices have been demonstrated, including high-efficiency photovoltaic cells [4], [5], high-speed electronic devices such as field-effect transistors [6], and optoelectronic devices like lasers [7], [8] and photodetectors [8], [9]. Nevertheless, the lattice mismatch between GaAs and Si presents a persistent bottleneck that must be addressed before fully harnessing the potential of GaAs/Si heterojunction-based devices.

To address the challenge of lattice mismatch, semiconductor grafting has emerged as a viable alternative [10]. This technique allows for the formation of abrupt heterointerfaces, regardless of lattice constant or thermal expansion coefficient differences. Such advancements are made possible by employing interface improvement strategies like chemical oxidation [11], [12], foreign ultrathin oxide deposition [13], [14], [15], [16], [17], and/or native oxide formation [18]. As a result, the interface density of states is greatly suppressed and the created heterojunction features a high quality that rivals conventional lattice-matched epitaxy [12], [16].

In this study, leveraging this grafting technique, we explore the fabrication and characteristics of GaAs/Si $p^+/n^+$ tunnel heterojunctions interfaced by an ultrathin oxygen-enriched layer (UOL). This interlayer was specifically engineered to improve interface quality and promote efficient carrier transport. The resulting tunnel diodes demonstrated negative differential resistance (NDR) at room temperatures with significantly improved peak-to-valley current ratios (PVCR), surpassing those achieved by direct bonding [19], [20] or epitaxy [21], [22], [23], [24]. These findings robustly demonstrate the dual functionality of the UOL as both a quantum tunneling medium and an interface passivation/improvement layer, highlighting its potential for

Jie Zhou, Yifan Wang, and Ziqian Yao contributed equally to this letter. The work is supported by a CRG grant (2022-CRG11-5079.2) by the King Abdullah University of Science and Technology (KAUST). The work also received partial support from DARPA H2 program under grant: HR0011-21-9-0109.

Jie Zhou, Yifan Wang Ziqian Yao, Jiarui Gong, Yang Liu, Yi Lu, Tsung-Han Tsai, Yiran Li, and Zhenqiang Ma are with the Department of Electrical and Computer Engineering, University of Wisconsin-Madison, Madison, WI 53706 United States. (e-mail: mazq@engr.wisc.edu).

Yara. S. Banda, Qingxiao Wang, Tien Khee Ng, and Boon S. Ooi are with Department of Electrical and Computer Engineering, King Abdullah University of Science and Technology, Thuwal 23955-6900, Saudi Arabia (e-mail: boon.ooi@kaust.edu.sa).

Carolina Adamo, Patrick Marshall, and Vincent Gambin are with Northrop Grumman Corporation, Redondo Beach, CA 90278, USA.

high-performance, lattice-mismatched heterojunction devices.

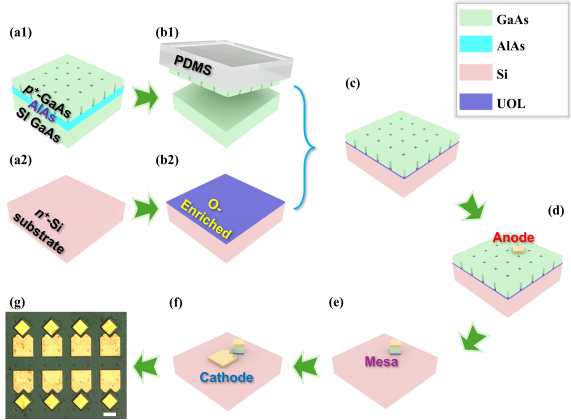

Fig. 1 Fabrication procedure illustration for the grafted GaAs/Si tunnel diodes. (a1) – (b1) Patterning, undercut, and retrieval of $p^+$-GaAs NM. (a2) – (b2) Preparation of the $n^+$-Si substrate by coating UOL. (c) Chemical thermal bonding of the $p^+$-GaAs NM to UOL-coated $n^+$-Si substrate. (d) Anode metallization on the $p^+$-GaAs NM, (e) mesa formation, and (f) cathode metallization on the exposed $n^+$-Si substrate. (g) Microscopic image of the completed GaAs/Si tunnel diode array. Scale bar: 50 μm.

## II. EXPERIMENT

The GaAs sample's epitaxial structure (Fig. 1 (a1)) involves a 300 nm p-type heavily doped GaAs ($4\times10^{19}$ cm$^{-3}$) cap layer, a 300 nm unintentionally doped (UID) AlAs sacrificial layer, and a semi-insulating (SI) GaAs substrate. The Si substrate (Fig. 1 (a2)) is *n*-type degenerately doped at $5\times10^{19}$ cm$^{-3}$. The GaAs epi and Si substrate first underwent a standard cleaning procedure consisting of sonication of samples in acetone, isopropyl alcohol, and deionized water. Subsequently, the GaAs epi was patterned with mesh holes, dry-etched to penetrate through GaAs cap, and selectively wet-etched to remove sacrificial AlAs layer. The cap GaAs was thus freestanding and retrieved from its host substrate using a polydimethylsiloxane (PDMS) (Fig. 1 (b1)), ready for the following procedures. Noteworthy, to improve the interface quality by enriching the oxygen content at the interface, a two-step oxidation was employed. The Si substrate (Fig. 1 (a2)) was first chemically oxidized in a piranha solution ($H_2SO_4$:$H_2O_2$ = 4:1) for 10 min [11], [12], [25], and then treated with five cycles of $Al_2O_3$ (~0.1 nm/cycle) at 200 °C by using trimethylaluminium (TMA) and $H_2O$ precursors. After these processes, an oxygen-enriched layer was formed on the Si surface (Fig. 1 (b2)). The released GaAs NM was transfer-printed to the UOL-coated Si substrate after removing the PDMS stamp. Later, to enable robust bonding, the GaAs/Si assembly was processed at 350 °C for 5 min within a rapid thermal annealer (Fig. 1 (c)). This process also enables an effective interface passivation on the GaAs side benefiting from ALD-$Al_2O_3$ [26], [27]. GaAs/Si tunnel diodes were then fabricated following a sequence of anode metallization (Au/Pt/Ti) on GaAs NM (Fig. 1 (d)), mesa formation and device isolation by dry etching the exposed GaAs material (Fig. 1 (e)), and cathode metallization (Au/Ti) on Si substrate (Fig. 1 (f)). The entire procedure was completed by an Ohmic metal annealing at 350 °C for 1 min, and a surface passivation with 80 cycles (~8 nm) of ALD-$Al_2O_3$ deposited at 200 °C using the same tool. The fabricated devices are presented in Fig. 1(g).

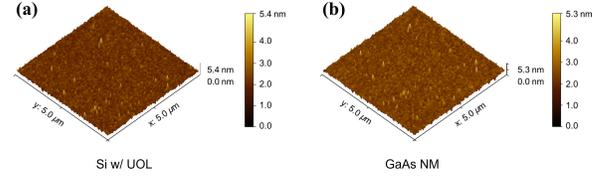

Fig. 2 AFM characterization of the synthesized $p^+$/$n^+$ GaAs/Si heterojunction. Surface morphology of (a) Si substrate with UOL, and (b) top surface of the grafted GaAs NM.

## III. RESULTS AND DISCUSSION

The surface morphology was characterized by AFM. Shown in Fig. 2(a), post UOL coating, the Si substrate exhibits an ultra-smooth surface, with a root-mean-square (RMS) of 0.361 nm. Similarly, the GaAs NM, even after undergoing transfer and thermal bonding processes, remained at an excellent roughness of 0.377 nm, seen in Fig. 2(b).

To examine the crystallinity of both grafted GaAs NM and the Si substrate and to directly correlate the junction properties with the device behaviors, STEM was conducted on a fabricated diode post electrical measurement. The low-magnification cross-sectional image of the entire junction, shown in Fig. 3(a), reveals the multi-layered heterostructure comprising the top metal contact (Au/Pt/Ti), grafted GaAs NM, interfacial UOL, and the Si substrate. The (wavy) interlayer thickness variation is due to the mismatched thermal expansion coefficients between Si and GaAs. It is noted that the interface $Al_2O_3$ has reflowed during annealing, leaving no voids at the interface. In Fig. 3(b), a high-resolution micrograph of the interface region is presented, with the Si lattice structure aligned along the [110] crystallographic axis. The amorphous interfacial UOL is measured to be ~1.5 to 2.5 nm in thickness. The single-crystalline nature of the grafted GaAs NM is illustrated in Fig. 3(c). EELS elemental mapping was also performed, as depicted in Fig. 3(d). The mapping clearly delineates the distribution of O, Ga, Si, Al, and C elements, as seen from Figs. 3(d i) to (d vi), respectively. The UOL can be distinctly identified at the heterointerface (Fig. 3(d ii)). This UOL serves to mitigate interface states and traps by chemically saturating the dangling bonds on adjacent GaAs and Si surfaces. Additionally, the sharp transition between the GaAs and Si layer, with visible interdiffusion of atoms such as Ga, As, and Si across the junction, further substantiates the role of the UOL as an effective physical diffusion barrier.

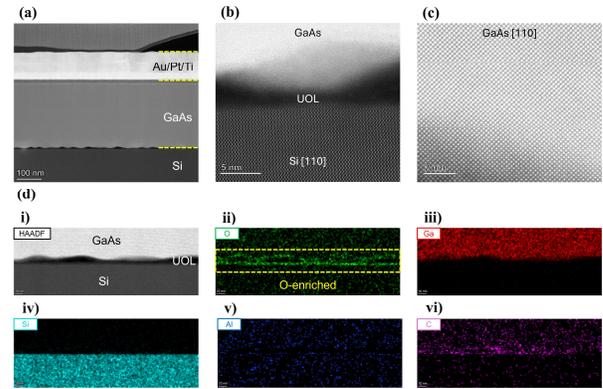

Fig. 3 Cross-sectional STEM micrographs and EELS mapping of the grafted GaAs/Si heterojunction. (a) An overview of the grafted heterojunction. (b) High-resolution STEM micrograph at the heterointerface, with the Si crystalline substrate along [110] orientation. (c) High-resolution STEM image of the grafted GaAs NM along [110] crystallographic orientation. (d) EELS elemental mapping of the heterojunction. An ultrathin oxygen-enriched interlayer is identified.

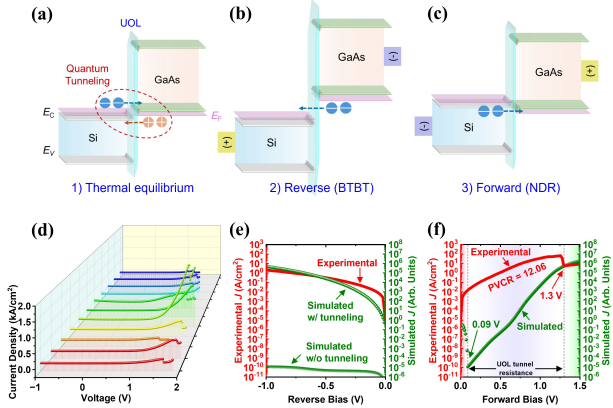

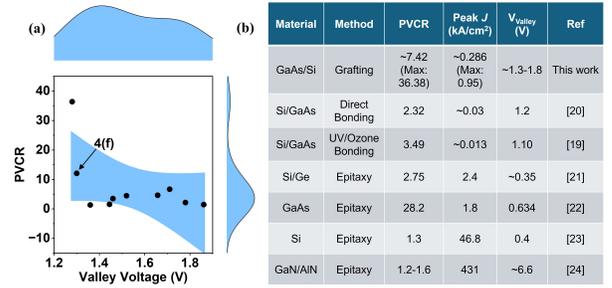

Fig. 4 Electrical characterization and simulation study on the grafted GaAs/Si $p^+/n^+$ tunnel diode. (a) Schematic illustration of the charge carrier tunneling process within the grafted GaAs/Si heterojunction interfaced by UOL. (b) and (c) Band alignment and charge transfer illustrations of the UOL-interfaced GaAs/Si tunnel diode under (b) reverse bias, and (c) forward bias. (d) Linear-scale measured I-V characteristics across 10 devices. (e) A typical I-V curve of the tunnel diode at reverse bias, resulted from band-to-band tunneling. (f) I-V curve at the forward bias, presenting obvious NDR behavior.

Besides its dual role in interface passivation and diffusion buffering, the UOL is engineered to be sufficiently thin to facilitate charge carrier transport across the heterointerface via quantum tunneling, as schematically depicted in Fig. 4(a). Given the degenerate doping of both GaAs and Si, the exponential rise in current under reverse bias (Fig. 4(e)) is attributed to band-to-band tunneling (BTBT), as depicted in Fig. 4(b). Under forward bias (Fig. 4(f)), negative differential resistance (NDR) is observed near a valley voltage of ~1.3 V, which is considerably higher than the theoretical/simulated value of ~0.09 V. This increased valley voltage is likely due to the presence of the nanometer-scale UOL, which introduces additional tunnel resistance for charge carriers. The tunneling characteristics of 10 devices are summarized in Fig. 4(d). A statistical analysis of these devices, presented in Fig. 5(a), shows the distribution pattern of PVCR and valley voltage, with further comparisons provided in the benchmark figure, Fig. 5(b). The measured metrics of our tunnel diodes include valley voltages ranging from 1.28 to 1.86 V, a peak tunnel current density of 0.95 kA/cm$^2$, and an average PVCR of 7.42 reaching a maximum of 36.38. These values represent a significant improvement over similar Si/GaAs tunnel diodes fabricated using other methods, such as direct bonding and UV/Ozone-assisted bonding [19], [20]. Additionally, when compared with previous epitaxial tunnel diodes from other material systems [21], [22], [23], [24], our PVCR is among the highest reported, highlighting the UOL's effectiveness in enhancing interface quality while imposing minimal adverse effects on device performance.

## IV. CONCLUSION

In this work, we successfully demonstrated the creation of GaAs/Si $p^+/n^+$ heterojunction tunnel diodes through the intentional incorporation of UOL. The tunnel diodes exhibited a high maximum PVCR of 36.38, underscoring the effectiveness of utilizing UOL as both a quantum tunneling medium and an interface improvement layer. This study also highlights the potential of incorporating such interlayers as a novel approach to creating lattice-mismatched tunnel diodes.

Fig. 5 Statistics and benchmark of our UOL-interfaced tunnel diodes. (a) Statistical PVCR and valley voltage obtained from 10 grafted tunnel diodes. (b) Benchmarking of our work against other types of tunnel diodes formed via direct bonding or epitaxy.

## REFERENCES

[1] H. Kroemer, "Nobel Lecture: Quasielectric fields and band offsets: teaching electrons new tricks," *Rev Mod Phys*, vol. 73, no. 3, pp. 783–793, Oct. 2001, doi: 10.1103/RevModPhys.73.783.

[2] Z. I. Alferov, "Nobel Lecture: The double heterostructure concept and its applications in physics, electronics, and technology," *Rev. Mod. Phys.*, vol. 73, no. 3, p. 767, 2001.

[3] H. Amano, "Nobel Lecture: Growth of GaN on sapphire via low-temperature deposited buffer layer and realization of $p$-type GaN by Mg doping followed by low-energy electron beam irradiation," *Rev. Mod. Phys.*, vol. 87, no. 4, pp. 1133–1138, Oct. 2015, doi: 10.1103/RevModPhys.87.1133.

[4] S. Kim, D.-M. Geum, M.-S. Park, C. Z. Kim, and W. J. Choi, "GaAs solar cell on Si substrate with good ohmic GaAs/Si interface by direct wafer bonding," *Sol. Energy Mater. Sol. Cells*, vol. 141, pp. 372–376, Oct. 2015, doi: 10.1016/j.solmat.2015.06.021.

[5] Y. Itoh, T. Nishioka, A. Yamamoto, and M. Yamaguchi, "GaAs heteroepitaxial growth on Si for solar cells," *Appl. Phys. Lett.*, vol. 52, no. 19, pp. 1617–1618, May 1988, doi: 10.1063/1.99058.

[6] G. M. Metze, H. K. Choi, and B. Tsaur, "Metal-semiconductor field-effect transistors fabricated in GaAs layers grown directly on Si substrates by molecular beam epitaxy," *Appl. Phys. Lett.*, vol. 45, no. 10, pp. 1107–1109, Nov. 1984, doi: 10.1063/1.95033.

[7] T. Wang, H. Liu, A. Lee, F. Pozzi, and A. Seeds, "1.3-μm InAs/GaAs quantum-dot lasers monolithically grown on Si substrates," *Opt. Express*, vol. 19, no. 12, pp. 11381–11386, Jun. 2011, doi: 10.1364/OE.19.011381.

[8] K. Tanabe, K. Watanabe, and Y. Arakawa, "III-V/Si hybrid photonic devices by direct fusion bonding," *Sci. Rep.*, vol. 2, no. 1, p. 349, Apr. 2012, doi: 10.1038/srep00349.

[9] C. Zeng, D. Fu, Y. Jin, and Y. Han, "Recent Progress in III–V Photodetectors Grown on Silicon," *Photonics*, vol. 10, no. 5, 2023, doi: 10.3390/photonics10050573.

[10] D. Liu et al., "Lattice-mismatched semiconductor heterostructures," *ArXiv Prepr. ArXiv181210225*, 2018.

[11] J. Gong et al., "Characteristics of grafted monocrystalline Si/β-Ga2O3 p–n heterojunction," *Appl. Phys. Lett.*, vol. 124, no. 26, p. 262101, Jun. 2024, doi: 10.1063/5.0208744.

[12] J. Gong et al., "Band alignment of grafted monocrystalline Si (001)/β-Ga2O3 (010) p-n heterojunction determined by X-ray photoelectron spectroscopy," *Appl. Surf. Sci.*, p. 159615, Feb. 2024, doi: 10.1016/j.apsusc.2024.159615.

[13] S. J. Cho et al., "P-type silicon as hole supplier for nitride-based UVC LEDs," *New J. Phys.*, vol. 21, no. 2, p. 023011, Feb. 2019, doi: 10.1088/1367-2630/ab0445.

[14] D. Liu et al., "229 nm UV LEDs on aluminum nitride single crystal substrates using p-type silicon for increased hole injection," *Appl. Phys. Lett.*, vol. 112, no. 8, p. 081101, Feb. 2018, doi: 10.1063/1.5011180.

[15] D. Liu *et al.*, "226 nm AlGaN/AlN UV LEDs using p-type Si for hole injection and UV reflection," *Appl. Phys. Lett.*, vol. 113, no. 1, p. 011111, Jul. 2018, doi: 10.1063/1.5038044.

[16] J. Zhou *et al.*, "Synthesis and characteristics of a monocrystalline GaAs/β-Ga2O3 p-n heterojunction," *Appl. Surf. Sci.*, vol. 663, p. 160176, Aug. 2024, doi: 10.1016/j.apsusc.2024.160176.

[17] J. Zhou *et al.*, "GaAs/GeSn/Ge n–i–p diodes and light emitting diodes formed via grafting," *J. Vac. Sci. Technol. B*, vol. 42, no. 4, p. 042213, Jul. 2024, doi: 10.1116/6.0003619.

[18] J. Zhou *et al.*, "Characteristics of Native Oxides-Interfaced GaAs/Ge np Diodes," *IEEE Electron Device Lett.*, vol. 45, no. 9, pp. 1669–1672, Sep. 2024, doi: 10.1109/LED.2024.3424461.

[19] K. Kim and J. Jang, "Improved Tunneling Property of p+Si Nanomembrane/n+GaAs Heterostructures through Ultraviolet/Ozone Interface Treatment," *Inorganics*, vol. 10, no. 12, 2022, doi: 10.3390/inorganics10120228.

[20] K. Kim, J. Jang, and H. Kim, "Negative differential resistance in Si/GaAs tunnel junction formed by single crystalline nanomembrane transfer method," *Results Phys.*, vol. 25, p. 104279, Jun. 2021, doi: 10.1016/j.rinp.2021.104279.

[21] W. Y. Fung, L. Chen, and W. Lu, "Esaki tunnel diodes based on vertical Si-Ge nanowire heterojunctions," *Appl. Phys. Lett.*, vol. 99, no. 9, p. 092108, Sep. 2011, doi: 10.1063/1.3633347.

[22] S. Ahmed, M. R. Melloch, E. S. Harmon, D. T. McInturff, and J. M. Woodall, "Use of nonstoichiometry to form GaAs tunnel junctions," *Appl. Phys. Lett.*, vol. 71, no. 25, pp. 3667–3669, Dec. 1997, doi: 10.1063/1.120475.

[23] M. W. Dashiell *et al.*, "Current-voltage characteristics of high current density silicon Esaki diodes grown by molecular beam epitaxy and the influence of thermal annealing," *IEEE Trans. Electron Devices*, vol. 47, no. 9, pp. 1707–1714, Sep. 2000, doi: 10.1109/16.861581.

[24] T. A. Growden *et al.*, "431 kA/cm2 peak tunneling current density in GaN/AlN resonant tunneling diodes," *Appl. Phys. Lett.*, vol. 112, no. 3, p. 033508, Jan. 2018, doi: 10.1063/1.5010794.

[25] H. Zhou, S. Alghmadi, M. Si, G. Qiu, and P. D. Ye, "Al2O3/ $\beta$ -Ga2O3(-201) Interface Improvement Through Piranha Pretreatment and Postdeposition Annealing," *IEEE Electron Device Lett.*, vol. 37, no. 11, pp. 1411–1414, Nov. 2016, doi: 10.1109/LED.2016.2609202.

[26] P. D. Ye *et al.*, "GaAs metal–oxide–semiconductor field-effect transistor with nanometer-thin dielectric grown by atomic layer deposition," *Appl. Phys. Lett.*, vol. 83, no. 1, pp. 180–182, Jul. 2003, doi: 10.1063/1.1590743.

[27] P. D. Ye *et al.*, "GaAs MOSFET with oxide gate dielectric grown by atomic layer deposition," *IEEE Electron Device Lett.*, vol. 24, no. 4, pp. 209–211, Apr. 2003, doi: 10.1109/LED.2003.812144.